\documentclass[aps,prl,
twocolumn, showpacs,
superscriptaddress,
amsfonts,amssymb,amsmath,
tightenlines,
twoside,
letterpaper,
bibnotes,
]{revtex4}

\usepackage{bm,color,mathrsfs,graphicx}

\begin{document}

\title{Electromagnon excitations in modulated multiferroics}

\author{A. Cano}
\email{cano@ill.fr}
\affiliation{Institut Laue-Langevin, 6 rue Jules Horowitz, B.P. 156, 38042 Grenoble, France}

\author{E. I. Kats}
\email{kats@ill.fr}
\affiliation {Institut Laue-Langevin, 6 rue Jules Horowitz, B.P. 156, 38042 Grenoble, France}
\affiliation {L. D. Landau Institute for Theoretical Physics, RAS, 117940 GSP-1, Moscow, Russia.} 

\date{\today}

\begin{abstract}
The phenomenological theory of ferroelectricity in spiral magnets presented in [M. Mostovoy, Phys. Rev. Lett. {\bf 96}, 067601 (2006)] is generalized to describe consistently states with both uniform and modulated-in-space ferroelectric polarizations. A key point in this description is the symmetric part of the magnetoelectric coupling since, although being irrelevant for the uniform component, it plays an essential role for the non-uniform part of the polarization. We illustrate this importance in generic examples of modulated magnetic systems: longitudinal and transverse spin-density wave states and planar cycloidal phase. We show that even in the cases with no uniform ferroelectricity induced, polarization correlation functions follow to the soft magnetic behavior of the system due to the magnetoelectric effect. Our results can be easily generalized for more complicated types of magnetic ordering, and the applications may concern various natural and artificial systems in condensed matter physics  (e.g., magnon properties could be extracted from dynamic dielectric response measurements).
\end{abstract}


\maketitle

The existence of magnetic materials that are also ferroelectrics
\cite{AI70,Smolenskii82,VG94} upsurge recurrent interest to study and to rationalize theoretically various models manifesting this possibility (see e.g. \cite{Katsura05,MO06,KH06,BG07}). The recent discovery of ferroelectricity in a family of the so-called frustrated magnets, see e.g. \cite{Kimura03}, is particularly instructive. In this case, ferroelectricity is a collateral effect of the magnetic ordering that, in contrast to earlier multiferroics, involves modulated structures \cite{DZ64,Fiebig05}. The current interest in this new class of ferroelectromagnets is twofold. 
On one hand the fundamental physics that result from the interplay between the 
electric and magnetic properties in these systems is not exhausted \cite{Fiebig05} and, on the other hand, many potential applications arise from this interplay \cite{SC07}.

The appearance of a modulated magnetic structure is accompanied by a distribution of 
polarization due to the inhomogeneous magnetoelectric effect \cite{Baryakhtar83}. 
This universal effect is similar to so-called flexoelectric effect in liquid crystals \cite{deGennes}. Microscopic studies have focused on the Dzyaloshinskii-Moriya 
interaction as the main source of the inhomogeneous magnetoelectric effect \cite{Katsura05}, 
and this idea seems also to be behind the phenomenological Landau-like approaches carried out until now in relation with ferroelectricity of magnetic origin \cite{MO06,Zhang07,Tewari07}. 
As a result, the main attention has been put to only antisymmetric parts of the effect (where the components of the magnetization enter in an antisymmetric combination). However, in a general case, the inhomogeneous magnetoelectric effect has also a non-vanishing symmetric part (see below). This is indeed the natural possibility from the phenomenological point of view \cite{Baryakhtar83,deGennes}, whatever the microscopic origin of this symmetric coupling. 

In this paper we show that this symmetric part of the inhomogeneous 
magnetoelectric effect is a key ingredient in the full description of materials 
exhibiting modulated magnetic structures. 
When it comes to the uniform polarization, indeed, only the antisymmetric part of the 
magnetoelectric effect plays a role in the agreement with previous theoretical 
publications (see e.g., \cite{Katsura05,MO06,Zhang07}, and more details below). 
However, to describe non-uniform, i.e. modulated in space, polarizations the magnetoelectric effect 
has to be considered in its full extent. This is precisely the case when addressing the so-called electromagnons, i.e., the normal modes characteristic of ferroelectromagnets that involve 
both polar modes and magnons in a hybridized way \cite{Smolenskii82,Baryakhtar70}.

Electromagnons were theoretically predicted long time ago considering 
uniformly ordered materials \cite{Baryakhtar70}. This consideration has been recently extended to the case of an antiferromagnet that becomes ferroelectric and then, through the inhomogeneous magnetoelectric effect, this induces an incommensurate magnetic structure \cite{deSousa07}. The first experimental evidences about electromagnons have been reported for materials in which, showing no instability towards ferroelectricity in the absence of magnetism, it is the appearance of a modulated magnetic structure what make them ferroelectrics \cite{Pimenov06,SL07}. 
Unfortunately, all theoretical studies performed so far for this latter case have been restricted to only antisymmetric parts of the magnetoelectric coupling \cite{Katsura07,Tewari07}. 
We show below that, contrary to what is implicitly assumed in these works, both symmetric and antisymmetric parts of the inhomogeneous magnetoelectric effect contribute nontrivially to the interplay between the fluctuations of magnetization and polarization in multiferroics. 

First of all, let us reconsider the uniform polarization that may arise due to the inhomogeneous 
magnetoelectric effect. To simplify our presentation and ease the algebra we neglect crystalline field anisotropy (a more realistic treatment does not affect our qualitative results). In this case, the magnetoelectric coupling term reads as
\begin{align}
F_{EM}& 
=- f_1 {\mathbf P} \cdot {\mathbf M}(\nabla \cdot {\mathbf M}) 
- f_2 {\mathbf P} \cdot [{\mathbf M}\times (\nabla \times {\mathbf M})],
\label{F_EM}\end{align}
where $\mathbf P$ is the polarization and $\mathbf M$ is the magnetization, and 
$f_{1}$ and $f_2$ are, in general, two different constants. The general form of this term is $-f_{kl,ij}P_kM_i\partial_lM_j$ \cite{Baryakhtar83}, 
where the magnetoelectric tensor reduces to 
$f_{kl,ij} = f_1 \delta_{ki}\delta_{lj} +f_2(\delta_{kl}\delta_{ij} - \delta_{kj}\delta_{li})$ 
in the isotropic case. 
Let us split the polarization into uniform $\overline {\mathbf P}$ and non-uniform (modulated) ${\mathbf P}'$  contributions: 
${\mathbf P}  = \overline{\mathbf P} + {\mathbf P}'$. 
From the magnetoelectric term, we then can extract the following contribution to the free energy of the system \cite{note0}:
\begin{align}
- f_{kl,ij} \overline{P}_{k}\negthinspace \int\negthinspace  d{\mathbf r} M_i\partial_lM_j = {\overline{P}_{k} \over 2}
\negthinspace \int \negthinspace d{\mathbf r} (f_{kl,ji} - f_{kl,ij} )M_i\partial_lM_j. 
\nonumber 
\end{align}
Thus, we see, the uniform polarization indeed couples to the magnetization only 
through the antisymmetric part of the magnetoelectric tensor. In the 
isotropic case that is
$f_{kl,ij} - f_{kl,ji} = 
(f_1 +f_2)(\delta_{ki}\delta_{lj} - \delta_{kj}\delta_{li} )$. 
So, as regards the uniform polarization, the relevant part of the magnetoelectric term \eqref{F_EM} can be written as 
$-{1\over 2}({f_1 +f_2 })\overline{\mathbf P}\cdot 
[{\mathbf M}(\nabla \cdot {\mathbf M})- ({\mathbf M}\cdot \nabla){\mathbf M}]
$, 
which coincides with the expression given in \cite{MO06}. 
However, it is important to keep in mind that there are actually two independent 
contributions, $f_1$ and $ f_2$, to the magnetoelectric coupling term. 

Let us now turn our attention to the magnetic features of the modulated magnets discussed in e.g. \cite{Fiebig05}. As we have mentioned, in these systems ferroelectricity is induced by magnetism and does not appears spontaneously by itself. In addition, the magnetoelectric effect is typically weak in comparison with pure magnetic contributions into the free energy. Therefore, in the first approximation, when determining the magnetic structure and dynamics it is possible to consider the magnetization separately. 

Most of these systems first undergo a transition from the magnetically disordered phase 
to a modulated structure similar to a longitudinal spin-density-wave (LSDW): 
\begin{align}
{\mathbf M} =(M_1\cos Q x,0,0),
\label{LSDW}\end{align}
and then to structure similar to a planar cycloid (PC):
\begin{align}
{\mathbf M} =(M_1\cos Q x,M_2\sin Q x,0).\label{cycloidal}
\end{align}
The dependence on $\mathbf P$ of Landau free energy density can be taken as
${1\over 2}\chi^{-1} P ^2 + F_{EM}$,
where $\chi$ is the dielectric susceptibility in the absence of magnetism \cite{note1} and $F_{EM}$ accounts for the magnetoelectric coupling. 
It can be easily seen that the inhomogeneous magnetoelectric effect \eqref{F_EM} implies the appearance of ferroelectricity as a result of the PC structure, but 
not of the LSDW \cite{MO06}. 

A simple way to reproduce this sequence of transitions 
is by considering Landau free energy density in the form
\begin{align}
F_M = {a\over 2 }M^2 + {b\over 4}M^4 + {1\over 2 }\sum _i M_i \hat L_{i}M_i,
\label{F_M}\end{align}
where the differential operators $\hat L_{i}$ describe the anisotropic  
softening that gives rise to the modulated structures. 
For small wavevectors 
one can approximate $\hat L_{i} = c\nabla^2$. The free energy \eqref{F_M} 
then is reduced to a familiar form (see e.g. \cite{Landau-Lifshitz_SP1}). Near the wavevector 
$\mathbf Q =(Q,0,0)$ of the modulated structures  
$a + \hat L_{i} \simeq \alpha_{i} + c_x\nabla^2_x + c\nabla^2_\perp$. 
Eq. \eqref{F_M} is then a natural generalization of the free energy considered in \cite{MO06,Zhang07} 
that provides the unified description of the spectrum 
at the relevant wavevectors (i.e., $\mathbf q = 0$ and $\pm \mathbf Q$). 
If $a,b,c > 0$, the system shows no 
instability with respect to uniform magnetizations.
Furthermore, if  $\alpha_x<\alpha_y<\alpha_z$, 
in the spirit of arguments presented in  \cite{MO06,Zhang07}, one can see that the free energy is minimized by the LSDW modulation in the parameter 
range $ \alpha_x < 0 $, $\alpha_{y} >0$, 
while the PC configuration occurs when $\alpha_x <0$ and $3\alpha_x \lesssim \alpha_y \lesssim \alpha_x/3$ (with $\alpha_z> 0 $ in both cases).

At this point, it is worth making the following comment on the label ``improper ferroelectrics'' that is frequently put to the systems of our interest (see e.g. \cite{Fiebig05}). Within above scenario $M_y$ or, more precisely, its $\pm {\mathbf Q}$ Fourier components,
can be seen as the order parameter of the LSDW-paraelectric $\leftrightarrow$ PC-ferroelectric transition. In the LSDW-paraelectric state Eq. \eqref{F_EM} has a term coupling $M_y$ and $P_y$ linearly so, in view of this linearity, the system should be labeled as proper rather than improper ferroelectric \cite{Strukov_Levanyuk}. This is in fact manifested in the corresponding anomalies \cite{MO06}.

As regards the dynamics, we are interested in the low-frequency excitations of the system. 
In the magnetically ordered phases, not too close to the transition points, 
these excitations are associated with variations of the magnetization in which its modulus is conserved. Thus the excitations are described by the Landau-Lifshitz equation \cite{Landau-Lifshitz_SP1}:
\begin{align}
\dot {\mathbf M} = \gamma {\mathbf M} \times {\mathbf H}_\text{eff},
\label{LL}\end{align}
where $\gamma$ is a constant and ${\mathbf H}_\text{eff} = (\hat L_x M_x,\hat L_y M_y,\hat L_z M_z)$.

The dispersion relations and correlation functions of our interest 
are obtained by linearizing this equation about the corresponding configurations of equilibrium. 
Then it is convenient to consider first of all the LSDW structure \eqref{LSDW}, and then a virtual transverse spin-density-wave (TSDW) modulation:
\begin{align}
{\mathbf M} =(0,M_2\sin Q x,0).\label{TSDW}
\end{align}
This would be the structure obtained from \eqref{F_M} if, contrary to what we assumed 
for the LSDW, $\alpha_y$ becomes negative with $\alpha_x >0$. 
This structure TSDW is of interest in its own right and besides, it turns out that the low-lying normal modes of the PC case can be found as the superposition of the normal modes associated with LSDW and TSDW structures. This notably simplifies the corresponding calculations. 

Skipping a large amount of tedious (although straightforward and simple) algebra, we come to the following results. They are obtained under the same assumptions as those for the static problem. That is, assuming that the anharmonicity is weak and that the spatial dispersion such that the generalized stiffness presents well defined minima only at the wavevectors $\mathbf q =0, \pm \mathbf Q$. This allows us to decouple the modes $\mathbf q $ and $\mathbf q \pm \mathbf Q$ from the rest \cite{Izyumov84}. To our purposes, higher-order satellites can be neglected.

As it could be expected, the low-lying modes of the LSDW represent independent oscillations of the magnetization along the $y$ and $z$ axes. The corresponding dispersion relations are such that
\begin{widetext}
\begin{align}
\omega^2_{y,z} (\mathbf q) =
\begin{cases}
2m_x^2 (\alpha_{z,y}-\alpha_x  + c_xq_x^2 + cq_\perp^2 )(-\alpha_x + cq^2 ), & q \to 0,\\
m_x^2 (\widetilde \alpha_{z,y} -2\alpha_x + cq'^2 )(\alpha_{y,z}-\alpha_x + c_xq_x'^2 + cq_\perp'^2 ), & {\mathbf q} = \mathbf Q + \mathbf q' \;(q'\to 0).\\
\end{cases}
\label{eigenfrecuencies}\end{align}\end{widetext}
Here $m_x =\gamma  M_1 /2$ and the quantities $\widetilde \alpha_i$ are defined such that $\hat L_i e^{\pm i2Qx} = \widetilde \alpha_i e^{\pm i2Qx }$. 
It is worth noticing that the closer the transition is, the smaller is the gap obtained at small wavevectors. Not only because of the smallness of $m_x$, what influences trivially on the whole spectrum, but also because $\alpha_x \to 0$. In fact, very close to the transition point $\omega_{y,z}(q \to 0)\sim q$. The gap obtained close to the wavevector of the modulation, on 
the contrary, is not so sensitive to the smallness of $\alpha_x$. 

Similar expressions are found for the TSDW structure \eqref{TSDW}. In this case, the low-lying modes are associated with oscillations of the magnetization along the $x$ and $z$ axes. Their dispersion relations can be obtained from \eqref{eigenfrecuencies} by replacing $y \leftrightarrow x$ in $m_x$, $\alpha_i$ and $\widetilde \alpha_i$. 

As we have mentioned, these results for the LSDW and TSDW structures merge in the PC case. 
Three low-lying normal modes are found such that the modulus of the magnetization is conserved in this latter structure. The modes associated with oscillations along the $x$ and $y$ axes coincide with that obtained previously for the TSDW and LSDW modulations respectively. Consequently, they have the same dispersion relations. 
The mode representing oscillations along the $z$ axis, however, is composed by the modes 
associated with the same type of oscillation in the TSDW and LSDW cases. In consequence, its 
dispersion relation can be expressed such that $\omega_{z,\text{PC}}^2({\mathbf q}) = \omega_{z,\text{LSDW}}^2({\mathbf q}) +\omega_{z,\text{TSDW}}^2({\mathbf q})$.

Let us now turn our attention to the fluctuations. Since the softness of the system is in its magnetic part, fluctuations of the magnetization $\delta \mathbf M$ will play the main role. 
Once Eq. \eqref{LL} is linearized, the corresponding correlation functions:
\begin{align}
M_{ij}({\mathbf q}) & \equiv \langle \delta M_i ({\mathbf q},\omega) \delta M_j(-{\mathbf q},-\omega) \rangle,
\label{correlationM}\end{align}
can be found with the aid of the fluctuation-dissipation theorem \cite{Landau-Lifshitz_SP1}. 
Within a first approximation, the polarization will follow these fluctuations in accordance with the magnetoelectric coupling \eqref{F_EM}. This results in fluctuations of the polarization $\delta \mathbf P$ of magnetic origin, i.e., electromagnons, whose correlation functions:
\begin{align}
P_{ij}({\mathbf q}) &\equiv \langle \delta P_i ({\mathbf q},\omega) \delta P_j(-{\mathbf q},-\omega) \rangle, 
\end{align}
can be expressed in terms of the previous quantities \eqref{correlationM}. 
The hybridization between magnons and polar modes reflects also in the form of cross correlation functions $\langle \delta P_i \delta M_j\rangle$. Within our approximation, these cross correlations can also be expressed in terms of the quantities \eqref{correlationM}. But since they essentially reveal the same information (see e.g. \cite{Ginzburg80}), we omit them in the following.

As regards magnetic correlation functions, the ``non-diagonal'' components
are zero for the structures of our interest ($M_{ij}=0$ if $i\not =j$). In addition, $M_{xx}=0$ in the LSDW case while $M_{yy} =0$ in the TSDW one. Close to the relevant wavevectors, the non-vanishing correlation functions are such that $M_{ii}^{-1}({\mathbf q}) \propto \tau\{[\omega^2 - \omega_{i}^2({\mathbf q})]^2 + \tau^{-2}\omega^2\}$, where $\tau$ is the (phenomenological) relaxation time of the corresponding oscillations. 

The fluctuations of the polarization of magnetic origin can be described in terms of the quantities $M_{ii}$. These fluctuations are such that the soft magnetic behavior at ${\mathbf q }\approx \pm {\mathbf Q}$ reflects in dipole polarization excitations with small wavevectors. And reversely, the soft magnetic response at $q \to 0$ reflects in polarization excitations with ${\mathbf q }\approx \pm {\mathbf Q}$, as can be seen in further formulas.
Note also that the magnetoelectric coefficients 
$f_1$ and $f_2$ enter separately in the following results.
That is, not only through the combination $f_1+f_2$ 
that determines the antisymmetric part of the magnetoelectric effect and, consequently, the space average of the polarization of the system.
(The $\mathbf q$-dependence associated with $\chi$ \cite{note1} is implicit in these formulas.) 

In the LSDW case we have 
\begin{align}
P_{xx}({\mathbf q})
& \sim f_1^2 \big\{
q^2_y \big[
M_{yy}({\mathbf q}-{\mathbf Q}) 
+
M_{yy}({\mathbf q}+{\mathbf Q})
\big]
\nonumber \\
&\qquad \;\;+
q^2_z
\big[
M_{zz}({\mathbf q}-{\mathbf Q}) 
+
M_{zz}({\mathbf q}+{\mathbf Q}) 
\big]
\big\}
,
\nonumber\\
P_{yy}({\mathbf q})&\sim 
g _{-2}^2(\mathbf q) 
M_{yy}({\mathbf q}-{\mathbf Q})
+
g _{+2}^2(\mathbf q)
M_{yy}({\mathbf q}+{\mathbf Q}),
\nonumber\\
P_{zz}({\mathbf q})&\sim 
g _{-2}^2(\mathbf q)
M_{zz}({\mathbf q}-{\mathbf Q})
+
g _{+2}^2(\mathbf q)
M_{zz}({\mathbf q}+{\mathbf Q}),
\nonumber\\
P_{xy}({\mathbf q})&\sim 
f_1q_y
\big[
g _{-2}(\mathbf q)  
M_{yy}({\mathbf q}- {\mathbf Q})
- 
g _{+2}(\mathbf q)
M_{yy}({\mathbf q}+{\mathbf Q}) 
\big],
\nonumber\\
P_{xz}({\mathbf q})&\sim 
f_1q_y
\big[
g _{-2}(\mathbf q)
M_{zz}({\mathbf q}- {\mathbf Q}) 
- 
g _{+2}(\mathbf q)
M_{zz}({\mathbf q}+{\mathbf Q}) 
\big],
\nonumber\\
P_{yz}({\mathbf q})&=0. \nonumber
\end{align}
where $g _{\pm2} (\mathbf q) = (f_1 +f_2)Q \pm f_2 q_x $. 
In the case of a pure symmetric coupling ($f_1 = -f_2 $), with no Dzyaloshinksii-Moriya-like contributions, all these correlation functions turn out to be $\propto q^2$. In consequence, the eventual importance of fluctuations of the polarization with small wavevectors is conditioned to the existence of a non-vanishing antisymmetric part in the magnetoelectric coupling. In any case, fluctuations with wavevectors ${\mathbf q}\approx \pm {\mathbf Q}$ are important irrespective to the above vanishing (they are entirely due to $f_1 \not =0$). These ``magnetic'' softenings of optical phonons in the LSDW paraelectric state can be used to measure the corresponding coefficients $f_1$ and $f_2$. This soft behavior in the absence of ferroelectricity has in fact been observed in \cite{SL07}.

Similarly, for the TSDW structure we find that
\begin{align}
P_{xx}({\mathbf q}) &\sim  f_2^2 
q_y^2 \big[
M_{xx}({\mathbf q}-{\mathbf Q} ) 
+
M_{xx}({\mathbf q}+{\mathbf Q})
\big]
,\nonumber\\
P_{yy}({\mathbf q}) &\sim  
g _{-1}^2(\mathbf q) 
M_{xx}({\mathbf q}-{\mathbf Q} ) 
+
g _{+1}^2(\mathbf q)
M_{xx}({\mathbf q}+{\mathbf Q})
\nonumber \\
&\quad +
f_1^2 q_z^2 \big[
M_{zz}({\mathbf q}-{\mathbf Q} ) 
+
M_{zz}({\mathbf q}+{\mathbf Q})
\big]
,\nonumber\\
P_{zz}({\mathbf q}) &\sim  
f_2^2 q_y^2 
\big[
M_{zz}({\mathbf q}-{\mathbf Q} ) 
+
M_{zz}({\mathbf q}+{\mathbf Q}) 
\big]
,\nonumber\\
P_{xy}({\mathbf q})  &\sim  
f_2 q_y
\big[
g _{-1}(\mathbf q)
M_{xx}({\mathbf q}-{\mathbf Q}) 
-
g _{+1}(\mathbf q)
M_{xx}({\mathbf q}+{\mathbf Q})
\big],\nonumber\\
P_{xz}({\mathbf q})  &=0,\nonumber\\
P_{yz}({\mathbf q}) &\sim  
-f_1 f_2 q_yq_z 
\big[
M_{zz}({\mathbf q}-{\mathbf Q}) 
+
M_{zz}({\mathbf q}+{\mathbf Q})
\big],\nonumber
\end{align}
where $g _{\pm1} (\mathbf q) = (f_1 +f_2)Q \pm f_1 q_x $. 
As in the previous case, the relevance of fluctuations of the polarization with small wavevectors is made conditional on the antisymmetric part of the magnetoelectric coupling while the relevance of fluctuations with wavevectros close to that of the modulated structure is not.

Finally, armed with the knowledge of correlation functions for LSDW and TSDW structures,
we are in the position to calculate the polarization correlation functions in the PC case as
$P_{ij}^\text{LSDW} + P_{ij}^\text{TSDW} + \Delta P_{ij}$. Here the first two terms represent the correlation functions obtained in for the LSDW and TSDW modulations respectively, and the third one is 
\begin{align}
\Delta P_{xx}({\mathbf q})&\sim  
q_x^2 \big[
f_1^2 \Pi_x ({\mathbf q},{\mathbf Q}) 
+
f_2^2 \Pi_y ({\mathbf q},{\mathbf Q}) 
\big],\nonumber \\
\Delta P_{yy}({\mathbf q}) &\sim  
q_y^2 \big[
f_2^2\Pi_x ({\mathbf q},{\mathbf Q}) 
+
f_1^2
\Pi_y ({\mathbf q},{\mathbf Q}) 
\big],\nonumber \\
\Delta P_{zz}({\mathbf q}) &\sim  
f_2^2q_z^2 \big[
\Pi_x ({\mathbf q},{\mathbf Q}) 
+
\Pi_y ({\mathbf q},{\mathbf Q}) 
\big]
,\nonumber \\
\Delta P_{xy}({\mathbf q})&\sim  
f_1f_2q_xq_y \big[
\Pi_x ({\mathbf q},{\mathbf Q}) 
+
\Pi_y ({\mathbf q},{\mathbf Q}) 
\big]
,\nonumber \\
\Delta P_{xz}({\mathbf q}) & \sim
f_1f_2q_xq_z \big[
\Pi_x ({\mathbf q},{\mathbf Q}) 
+
\Pi_y ({\mathbf q},{\mathbf Q}) 
\big]
,\nonumber \\
\Delta P_{yz}({\mathbf q} )&\sim  
f_1f_2q_yq_z \big[
\Pi_x ({\mathbf q},{\mathbf Q}) 
+
\Pi_y ({\mathbf q},{\mathbf Q}) 
\big]
.\nonumber 
\end{align}
where $\Pi_i ({\mathbf q},{\mathbf Q}) =m_i^2 [M_{ii}( {\mathbf q}-{\mathbf Q}) + M_{ii}({\mathbf q}+{\mathbf Q})]$. 
These additional contributions are relevant for the fluctuations with wavevectors close to $\pm \mathbf Q$. 
It is worth noticing that, within our approximations, there are no contributions $\propto m_1 m_2$ to these correlations functions.

In summary, we have presented the key aspects of the phenomenological theory of ferroelectricity 
of magnetic origin that allow one to treat consistently, on the same footing, uniform and non-uniform polarizations. One of its key points is the accounting for all the possible contributions, symmetric and antisymmetric ones, to the inhomogeneous magnetoelectric effect. 
Considering the most representative modulated magnetic structures, we have employed this theory to study the corresponding fluctuations of magnetization and polarization. Our main findings are the following. 
i) The aforementioned fluctuations turn out to be interdependent as long as magnetic order appears, irrespective of whether this order induces an uniform polarization or not. This implies that ferroelectricity may not be strictly necessary when addressing applications based on such an interplay. 
ii) The softness of the system reveals in the form of large fluctuations at both, small wavevectors and wavevectors close to that of the modulated structure. At small wavevectors, the fluctuations of the polarization of magnetic origin can be associated with the antisymmetric part of the magnetoelectric effect (that eventually gives rise to the uniform polarization of the system). At wavevectors close to that of the modulated structure, however, the situation is different. In this case, the fluctuations are due to both symmetric and antisymmetric parts of the effect. Consequently, in view of the relativistic nature of the antisymmetric part, one can expect here a leading role of the symmetric contribution (which has its origin in the exchange interaction).

We acknowledge very fruitful discussions with L.P. Regnault, I.E. Dzyaloshinskii and A.P. Levanyuk.
A.C. was supported by a postdoctoral fellowship from Fundaci\' on Ram\' on Areces.


\begin{references}
\bibitem{note0} To derive the latter expression we transform the magnetoelectric tensor into $(f_{kl,ij} + f_{kl,ij} )/2$ and integrate by parts one of the resulting terms.
\bibitem{note1} Strictly speaking, in the absence of magnetism we should write ${1\over 2}(\chi^{-1}_0 P^2 - {\mathbf E}\cdot{\mathbf P})$, where $\chi_0$ is a constant and the field $\mathbf E$ is caused by inhomogeneous distributions of ${\mathbf P}$. That is, the electric field obtained from the equations of electrostatics $\nabla \times {\mathbf E} =0$ and $\nabla \cdot({\mathbf E} +4\pi {\mathbf P})=0$. 
This results into the susceptibility $\chi$.
\end{references}
\end{document}